\documentstyle[epsf,11pt]{article}
\begin{document}

\title{
Reducing quasi-ergodicity in a double well potential by Tsallis Monte
Carlo simulation
}

\author{
Masao Iwamatsu$^{\dagger}$\thanks{Permanent address. 
E-mail:iwamatsu@ce.hiroshima-cu.ac.jp} and Yutaka Okabe$^{\dagger}$
\\$^{*}$Department of Computer Engineering, Hiroshima City University\\ 
Hiroshima 731-3194, Japan\\ and\\
$^{\dagger}$Department of Physics, Tokyo Metropolitan University\\ 
Hachioji, Tokyo 192-0397, Japan }

\date
{
}

\maketitle

\begin{abstract}
A new Monte Carlo scheme based on the system of Tsallis's generalized 
statistical mechanics is applied to a simple double well potential 
to calculate the canonical thermal average of potential energy. 
Although we observed serious quasi-ergodicity when using the standard 
Metropolis Monte Carlo algorithm, this problem is largely reduced by 
the use of the new Monte Carlo algorithm.  Therefore the ergodicity 
is guaranteed even for short Monte Carlo steps if we use this new 
canonical Monte Carlo scheme. \\

\begin{flushleft}



\end{flushleft}
\end{abstract}

\newpage
\section{Introduction}

The ergodic hypothesis is fundamental to statistical mechanics. 
This hypothesis states that the time average of an observable 
event equals the phase-space average. In practical application,
however, problems can arise in various types of simulations 
if the system must overcome high energy barriers to reach other 
regions of phase space. In that case, the length of a simulation 
needed in order to obtain enough statistical samples of all 
regions of phase space may be extremely long. In the Monte Carlo 
simulation, the errors arise as a consequence of the finite 
length of the Monte Carlo walk. This error can be serious in 
canonical Monte Carlo sampling, especially at low temperatures.
This problem, referred as "quasi-ergodicity" by Valleau and 
Whittington~\cite{VW}, appears even in the simplest double well 
problem, where the two wells are separated by a large 
barrier~\cite{FFD}.

Recently several authors~\cite{Pe,TS,AS1} have proposed a 
generalized simulated annealing method to locate the global 
minimum based on the generalized statistical mechanics 
proposed by Tsallis~\cite{Ts}, which can overcome the slow 
convergence of the traditional simulated annealing
method~\cite{KGV} based on the standard Monte Carlo scheme. 
Subsequently, Andricioaei and Straub have pointed out that 
their generalized Monte Carlo (Tsallis Monte Carlo)~\cite{AS2} 
method may be used to overcome the quasi-ergodicity appearing 
in the canonical average at constant temperatures.

In this short note we use a new Monte Carlo scheme~\cite{AS2,AS3,SA} 
based on Tsallis's generalized statistical mechanics~\cite{TS} 
for the calculation of the thermal average of the potential energy 
of a simple double well potential at constant temperatures.
We examine the performance of this new algorithm designed to 
overcome the quasi-ergodicity.  

\section{A generalized Monte Carlo scheme} 

In the generalized statistical mechanics proposed by 
Tsallis~\cite{Ts}, a crucial role is played by the generalized 
entropy $S_{q}$ defined as 
\begin{equation}
S_{q}=k\frac{1-\sum p_{i}^{q}}{q-1}
\label{eq:1}
\end{equation}
where $q$ is a real number which characterizes the statistical 
mechanics, and $p_{i}$ is the probability of states $i$. This 
entropy $S_{q}$ becomes the usual Gibbs-Shannon entropy
$S_{1}=-k\sum p_{i}\ln p_{i}$ when $q\rightarrow 1$. By 
maximizing the generalized entropy with the constraints 
\begin{equation}
\sum p_{i}=1\;\;\;\mbox{and}\;\; \sum p_{i}^{q}\epsilon_{i}=\mbox{const} 
\label{eq:2}
\end{equation}
where $\epsilon_{i}$ is the energy spectrum, the statistical weight 
for the generalized canonical ensemble characterized by the 
parameter $q$ is given by the Tsallis weight
\begin{equation}
p_{i}^{q}=\exp\left(-\overline{\beta}\overline{\epsilon}_{i}\right)/Z_{q}^{q} 
\label{eq:3}
\end{equation}
with
\begin{equation}
\overline{\epsilon}_{i}=\frac{q}{\overline{\beta}(q-1)}
\ln\left[1-(1-q)\overline{\beta}\epsilon_{i }\right]
\label{eq:4}
\end{equation}
where $\overline{\epsilon}_{i}$ is the generalized energy and 
$\overline{\beta}$ is the Lagrange multiplier, which plays the
role of the generalized temperature in the Tsallis statistical
mechanics.  We should note here that the statistical weight is 
{\it not} given by the probability distribution $p_{i}$ {\it but} 
rather by $p_{i}^{q}$ in order to calculate thermal averages, as 
can be seen in (\ref{eq:2}). The generalized canonical partition 
function $Z_{q}$ in (\ref{eq:3}) is given by
\begin{equation}
Z_{q}=\sum \exp\left(-\overline{\beta}\overline{\epsilon}_{i}/q\right) 
\label{eq:5}
\end{equation}
In the limit $q\rightarrow 1$, $\overline{\epsilon}_{i}\rightarrow
\epsilon_{i}$
and $\overline{\beta}\rightarrow\beta=1/kT$ where $T$ is the usual 
temperature~\cite{CT}, then the Tsallis weight (\ref{eq:3}) becomes the 
Boltamann weight $p_{i}=\exp(-\beta\epsilon_{i})/Z_{1}$ of the usual 
canonical ensemble.

In this generalized statistical mechanics, Andricioaei and 
Straub~\cite{AS1} have noted that since the average of an 
observable $O$ is defined by 
\begin{equation}
<O>_{q}=\sum p_{i}^{q}O_{i},
\label{eq:6}
\end{equation}
the detailed balance condition should be written as 
\begin{equation}
p_{i}^{q}W_{ij}=p_{j}^{q}W_{ji}
\label{equation}
\end{equation}
where $W_{ij}$ is the element of the transition matrix. Based 
on this observation, they~\cite{AS1} pointed out that the 
generalized Monte Carlo algorithm~\cite{TS} originally proposed 
by Penna~\cite{Pe} in his generalized simulated annealing method 
cannot satisfy the detailed balance.  They~\cite{AS1} then 
proposed a new generalized Monte Carlo scheme, where the 
acceptance probability $p$ of the Monte Carlo move is given by
\begin{equation}
p=\min\left[1,\exp(-\overline{\beta}\Delta \overline{\epsilon})\right] 
\label{eq:7}
\end{equation}
which has a similar form to the familiar Metropolis 
algorithm~\cite{MRRTT} 
\begin{equation}
p=\min\left[1,\exp(-\beta\Delta \epsilon)\right] \label{eq:8}
\end{equation}
where $\Delta\overline{\epsilon}$ is the increase of the 
generalized energy (\ref{eq:4}) while $\Delta\epsilon$ is that 
of the usual energy. Andricioaei and Straub~\cite{AS1} combined 
this Monte Carlo algorithm with the usual simulated annealing 
technique~\cite{KGV} to optimize the conformation of tetrapeptides.
They showed that their algorithm is more effective than the 
standard simulated annealing based on the usual molecular
dynamics or Monte Carlo methods.

Recently, Andricioaei and Straub~\cite{AS2,AS3} pointed out that their 
generalized Monte Carlo scheme~\cite{AS1} for simulated annealing can 
also be used to calculate the standard canonical ensemble averages at 
constant temperatures
\begin{equation}
<O>_{1}=\frac{\sum O_{i}\exp(-\beta \epsilon_{i})}
{\sum \exp(-\beta \epsilon_{i})}
\label{eq:9}
\end{equation}
using the general rule of the importance sampling~\cite{VW}
\begin{equation}
<f>=\sum f_{i}=\sum \left(\frac{f_{i}}{g_{i}}\right)g_{i}
\label{eq:8b}
\end{equation}
Then, equation (\ref{eq:9}) can be calculated using the generalized
Monte Carlo scheme from
\begin{eqnarray}
<O>_{1}&=&\frac{ \sum
O_{i}\exp\left(-\left(\beta\epsilon_{i}-\overline{\beta}
\overline{\epsilon}_{i}
\right)\right)\exp\left(-\overline{\beta}
\overline{\epsilon}_{i}\right) }{\sum
\exp\left(-\left(\beta\epsilon_{i}-\overline{\beta}
\overline{\epsilon}_{i}\right) 
\right)\exp(-\overline{\beta}\overline{\epsilon}_{i})} 
\label{eq:10}\\
&=& \frac{< O \exp\left(-\left(\beta\epsilon-
\overline{\beta}\overline{\epsilon} 
\right)\right)>_{q}}{<\exp\left(-\left(\beta\epsilon-
\overline{\beta}\overline{\epsilon}\right)\right)>_{q}}
\label{eq:11}
\end{eqnarray}
where the generalized average $<>_{q}$ is defined by (\ref{eq:6}). 
In practice, the equilibrium thermal average in the standard 
canonical ensemble can be calculated by conducting these 
"generalized" Monte Carlo steps and accumulating the samples by
\begin{equation}
<O>_{1}=\frac{\sum O_{j} \exp\left(-\left(\beta\epsilon_{j}- 
\overline{\beta}\overline{\epsilon}_{j}\right)\right)}
{\sum \exp\left(-\left(\beta\epsilon_{j}-
\overline{\beta}\overline{\epsilon}_{j}\right)\right)}
\label{eq:12}
\end{equation}
where the sum $j$ runs over to the generalized Tsallis Monte 
Carlo steps produced by eq.(\ref{eq:7}) instead of the usual 
Metropolis algorithm (\ref{eq:8}). The equation (\ref{eq:12}) 
is a generalized expression, and we may choose 
$\overline{\beta}=\beta$.  Then, the equation (\ref{eq:12}) 
reduces to
\begin{equation}
<O>_{1}=\frac{\sum O_{j} \exp\left(-\beta\left(\epsilon_{j}- 
\overline{\epsilon}_{j}\right)\right)}
{\sum \exp\left(-\beta\left(\epsilon_{j}-
\overline{\epsilon}_{j}\right)\right)},
\label{eq:12x}
\end{equation}
which has been used in \cite{AS2,AS3}.  From now on, the actual 
calculation will be done by choosing $\overline{\beta}=\beta$.

We would like to point out here that a similar reweighting 
technique is used in the histogram method~\cite{FS} and 
the multicanonical ensemble method~\cite{BN}. Andricioaei and 
Straub applied this Tsallis Monte Carlo algorithm to the 
two-dimensional Ising system~\cite{AS2}, a classical one-dimensional 
potential and a 13-atom Lennard-Jones cluster~\cite{AS3}, and 
showed that this new Monte Carlo algorithm is more effective 
than the standard Metropolis algorithm.  They also proposed the
generalized molecular dynamics based on Tsallis generalized statistical
mechanics~\cite{AS3,SA}

\section{Application to a classical double well potential} 

In order to check the performance of this new Monte Carlo 
algorithm~\cite{AS2,AS3} in a classical system such as molecules 
and liquids, we look at the problem of a classical particle 
in a one-dimensional double well potential defined by
\begin{equation}
V(x)=\delta\left(3 x^{4}+4(\alpha-1)x^{3}-6\alpha x^{2}\right)+1 
\label{eq:13}
\end{equation}
where
\begin{equation}
\delta=\frac{1}{2\alpha+1}
\label{eq:14}
\end{equation}
which was examined by Frantz {\it et al.}~\cite{FFD} to demonstrate 
the quasi-ergodicity of the standard Metropolis algorithm. The 
potential given by (\ref{eq:13}) has the double minimum at $x=1$ 
and $x=-\alpha$; therefore, it represents a symmetrical double well 
when $\alpha=1$ and a single well when $\alpha=0$. This potential 
is more conveniently characterized by the parameter $\gamma$~\cite{FFD} 
\begin{equation}
\gamma=\frac{V(0)-V(-\alpha)}{V(0)-V(1)}
=\alpha^{3}\left(\frac{\alpha+2}{2\alpha+1}\right)
\label{eq:15}
\end{equation}
In figure 1, we show the asymmetric double well potential $V(x)$ 
when $\gamma=0.9$ as well as the equilibrium Tsallis weight 
$p_{\rm T}\propto \exp(-\beta \overline{V}(x))$, which is renormalized, 
and Boltzmann weight $p_{\rm B}=\exp(-\beta V(x))/Z_{1}$. It is obvious 
that the statistical weight $p_{\rm T}$ of the Tsallis generalized 
statistical mechanics has a greater chance of crossing the barrier 
and, therefore, a greater possibility of avoiding quasi-ergodicity.

Alternatively this Tsallis's statistical mechanics can be regarded 
as a search-space smoothing method which deforms the rugged potential 
landscape by a smoother one.  In figure 2, we show the smoothed 
potential $\overline{V}(x)$ calculated from (\ref{eq:4}), which has 
a smoother landscape and a lower energy barrier than the  
original landscape $V(x)$.

We use this double well potential to look at the quasi-ergodicity 
of the standard as well as the generalized Monte Carlo scheme by 
examining the classical average potential energy~\cite{FFD} 
\begin{equation}
<V>=\frac{\int V(x)\exp\left(-\beta V(x)\right)dx}
{\int \exp\left(-\beta V(x)\right)dx }
\label{eq:16}
\end{equation}
which can be calculated rigorously by numerical quadrature as a 
function of the temperature $\beta$ and the asymmetry parameter 
$\gamma$. We employ the same Monte Carlo procedure of Frantz 
{\it et al.}~\cite{FFD} and use the uniform sampling distribution 
from previous position $x$ to new position $x^{'}$ given by
\begin{eqnarray}
T(x^{'}|x)&=&\frac{1}{\Delta}\;\;\;\mbox{for}
\;\;|x-x^{'}|<\frac{\Delta}{2}\ \nonumber \\
&=& 0\;\;\;\mbox{otherwise}
\label{eq:17}
\end{eqnarray}
The step size $\Delta$ can be chosen to be large enough for the 
one-dimensional problem so that the quasi-ergodicity can
be reduced~\cite{FFD}.  However, we follow Frantz {\it et
al.}~\cite{FFD} and use the scaling $\Delta=2.5/\sqrt{\beta}$, 
which maintains an approximately 50\% acceptance rate at all 
temperatures $\beta$, by considering the application to 
multidimensional systems.  This 50\% acceptance is commonly 
employed in the classical Monte Carlo community~\cite{AT}.

Since the error caused by the quasi-ergodicity depends upon 
the walk initialization, we start the Monte Carlo walk at 
the global minimum $x=1$ and at the metastable minimum $x=-\alpha$. 
If the walk starts at the global minimum, then the average 
potential $<V>$ will be low if the walk is quasi-ergodic, as 
the distribution associated with the higher energy well around 
$\alpha$-minimum will be insufficiently sampled. While the walk 
is initialized at $-\alpha$, the walk may be trapped at this 
metastable $\alpha$-well and the average potential $<V>$ 
will be too high.

In figure 3, we show the average potential energy $<V>$ calculated 
from (i) the standard Metropolis Monte Carlo and (ii$\sim$iv) the 
generalized Tsallis Monte Carlo algorithm as a function of
temperature $\beta$. Similar curves can be found in figure 2 of 
reference \cite{FFD}; however, those authors plotted the results 
from the usual Metropolis algorithm initialized at the global 
minimum $x=1$ and at the random positions. In order to check 
the quasi-ergodicity of the walk, we instead show two curves 
initialized at the global minimum $x=1$ and the metastable 
$\alpha$-minimum at $x=-\alpha$ for each scheme. The average 
potential $<V>$ is obtained from 100 independently initialized 
walks, each consisting of 500 warm-up steps followed by $10^{4}$ 
steps with data accumulation.  The average energy $<V>$ calculated 
from the standard Metropolis algorithm shows large errors due to 
the quasi-ergodicity. In particular, the walk initialized at 
the metastable $\alpha$-minimum greatly overestimates the potential 
energy. The walk starting from the global minimum also shows 
noticeable error and underestimates the average energy, as expected.

In contrast to the results from the standard Metropolis algorithm, 
the results from the generalized Monte Carlo algorithm~\cite{AS2,AS3} 
show excellent agreement with the numerically exact result. The 
walks started from the global minimum $x=1$ produce the average 
energy which is almost indistinguishable from the exact result up 
to the lowest temperature (high $\beta$) examined. Even the walk 
started from the metastable $\alpha$ minimum produces the energies 
that are fairly accurate up to rather low temperature $\beta \sim 15$ 
for $q=1.5$, which shows that this generalized algorithm does 
a much better job than the standard Monte Carlo scheme. The agreement 
with the exact result is further improved if we increase the value 
of parameter $q$ from $1.5$ (see Figure 3).  It should be noted 
that the probability distribution associated with the parameter 
$q=2$ is given by the Cauchy distribution and with $q>5/3$ by the 
super-diffusive L\'evy distribution~\cite{TS}. The theoretical upper 
boundary is $q=3$. The quasi-ergodicity can be largely circumvented 
by the use of Tsallis's generalized statistical mechanics with
$q>1$, as shown in figure 3. 

In order to overcome the quasi-ergodicity, the Monte Carlo
walk has to pass many times through the energy barrier.
Figure 4 compares the time series of the Monte Carlo walk
generated by the standard Monte Carlo scheme ($q=1$) and that 
of the generalized Monte Carlo scheme.  Similar diagrams
can be found in reference ~\cite{AS3}.  Although the 
trajectory of the standard Monte Carlo scheme is mostly 
confined within the metastable $\alpha$ minimum, the number 
of barrier crossing of the Tsallis Monte Carlo scheme is 
significantly greater.  This barrier crossing occurs more 
often when the parameter $q$ is larger.  The quasi-ergodicity 
can be removed by Tsallis's generalized Monte Carlo scheme 
by increasing the magnitude of the parameter $q$ from $1$.

In figure 5, we show the average energy $<V>$ as a function of 
the asymmetry parameter $\gamma$. Again a similar diagram can 
be found in reference~\cite{FFD}. Here again we can
clearly observe a large error for the initialization at the 
metastable $\alpha$ minimum when the standard Monte Carlo
procedure is used. In this case, the superiority of Tsallis's 
generalized Monte Carlo scheme over the traditional one is
also obvious.

In table 1 we showed the effect of quasi-ergodicity in the 
convergence when $\gamma=0.9$ and $\beta=10$ for both 
the standard Monte Carlo and the generalized Monte Carlo 
algorithms with $q=1.5$, each originating at the metastable 
$\alpha$-minimum. We show the average energy $<V>$ and the 
standard deviation (STD) obtained from 100 independently 
initialized walks. We found that the problem of quasi-ergodicity 
obtained with the standard Metropolis algorithm is still present, 
even if we increase the number of Monte Carol steps. On the other 
hand the result from the generalized Monte Carlo simulation is 
significantly improved by increasing the number of Monte Carlo
steps, and ergodicity can be recovered.  Furthermore, STD
decreases as $\sim 1/\sqrt{N}$ when the Monte Carlo step $N$ is 
increased for the generalized Monte Carlo simulation, while such 
a systemic decrease is not observed for the standard Monte 
Carlo simulation.

This convergence can be clearly visualized by plotting the
visiting probability $P(x)$ of the Monte Carlo walks, which 
should be given by the Boltzmann weight for the standard
Monte Carlo scheme and by the Tsallis weight (\ref{eq:3}) for
the generalized Monte Carlo scheme.  In figure 6, we compare 
$P(x)$ generated by the standard and generalized Monte Carlo 
walks with the Boltzmann and the Tsallis weights.  The visiting 
probability $P(x)$ converges rapidly to the Tsallis weight as 
the number of the Mone Carlo walks is increased for the 
generalized Monte Carlo scheme.  On the other hand, $P(x)$ 
recovers the Boltzmann weight only when the Monte Carlo steps 
are increased upto $\sim 10^{6}$ in the standard Monte 
Carlo scheme.

It is also interesting to discuss the convergence in the context of
the "generalized ergodic measure" of Thirumalai, Mountain, and
Kirkpatrick~\cite{TMK,AS3}. Using the energy metric, the ergodic measure
$d_{V}(n)$ is defined by
\begin{equation}
d_{V}(n)=\left(V^{a}(n)-V^{b}(n)\right)^{2},
\label{eq:18}
\end{equation}
where $V^{a}(n)$ and $V^{b}(n)$ are the move average of the potential
energy $V$ along the trajectories $a$ and $b$, which
are defined from (\ref{eq:12x}) by
\begin{eqnarray}
V^{a}(n)&=&\frac{\sum_{j=0}^{n} V_{j}^{a} \exp\left(-\beta\left(V_{j}^{a}- 
\overline{V}_{j}^{a}\right)\right)}
{\sum_{j=0}^{n} \exp\left(-\beta\left(V_{j}^{a}-
\overline{V}_{j}^{a}\right)\right)}.
\label{eq:19}
\end{eqnarray}
for the generalized Tsallis Monte Carlo scheme, and
\begin{eqnarray}
V^{a}(n)&=&\frac{\sum_{j=0}^{n} V_{j}^{a} \exp\left(-\beta V_{j}^{a}\right)}
{\sum_{j=0}^{n}1}.
\label{eq:19x}
\end{eqnarray}
for the standard Monte Carlo scheme.
For an ergodic system, Thirumalai {\it et al}.~\cite{TMK} suggested that
the ergodic measure converges as $d_{V}(n)\rightarrow \infty$ if
$n\rightarrow \infty$.  They found that
\begin{equation}
d_{V}(n)\simeq d_{V}(0)\frac{1}{D_{V}n}
\label{eq:20}
\end{equation}
where the "diffusion" constant $D_{V}$ depends on temperature.  In figure
7, we show $d_{V}(0)/d_{V}(n)$ for $\gamma=0.9$ potential obtained from 100
independent pairs ($a$, $b$) of Monte Carlo walks as a function of Monte Carlo
steps $n$.  We chose the trajectory starting from the metastable minimum
$x=-\alpha$  (=-0.9163) and that from the stable minimum $x=1$,
respectively, as the two independent trajectories $a$ and $b$.  It is
clear from figure 7 that the normalized inverse of the ergodic measure
$d_{V}(0)/d_{V}(n)$ grows linearly with the Monte Carlo steps $n$.
Therefore, the ergodic measure $d_{V}(n)$ decreases according to
(\ref{eq:20}), and the diffusion constant $D_{V}$ can be determined.

Figure 8 shows the diffusion constant $D_{V}$ as a function of temperature
$\beta$.  The diffusion constant is a decreasing function of the inverse
temperature $\beta$; it is more difficult to recover the ergodicity when
the temperature is lowered.  However, at lower temperatures $\beta>20$ with
$\beta\Delta V>1$ where $\Delta V=V(-\alpha)-V(1)=0.1$, the diffusion
constant starts to increase again for the Tsallis Monte Carlo walks ($q\neq
1$).  This is due to the fact that the equilibrium thermal distribution
around the metastable minimum at $x=-\alpha$ becomes negligibly small at
such low temperatures, and the ergodicity of the trajectory among the
stable and the metastable minima is
irrelevant once trajectories $a$ and $b$ both fall into the stable
minimum around $x=1$.

It is apparent from figure 8 that the diffusion constant $D_{V}$ obtained
from the standard Monte Carlo algorithm obeys the usual activation form
\begin{equation}
D_{V} \sim \exp\left(-\beta E_{b}\right)
\label{eq:21}
\end{equation}
where the activation energy $E_{b}$ is estimated to be $E_{b}\simeq 0.4$,
which is the same order of magnitude as the barrier height ($\sim 1$)
between two wells, as shown in figure 1.   On the other hand, the
diffusion
constant $D_{V}$ obtained from the Tsallis generalized Monte Carlo scheme is
characterized by a power-law form
\begin{equation}
D_{V} \sim \beta^{-\eta}
\label{eq:22}
\end{equation}
with the exponent $\eta$.  This result is conceivable, based on the
acceptance
probability (\ref{eq:7}) using the Tsallis weight, because
\begin{eqnarray}
D_{V} &\sim& \exp(-\beta\bar{E}_{b})
\nonumber \\
&\sim& \left[1-(1-q)\beta E_{b} \right]^{-q/(q-1)}
\label{eq:23}
\end{eqnarray}
and $D_{V}\sim \beta^{-q/(q-1)}$.  We note in figure 8 that $\eta\sim 1.6$
for $q=1.5$ and $\eta\sim 0.85$
for $q=2.5$.  These values are close to $q/(q-1)=3$ for $q=1.5$ and
$q/(q-1)=1.66$ if $q=2.5$.  Therefore, the diffusion constant decreases
more slowly as a function of the inverse temperature $\beta$ as
$\beta^{-\eta}$ for the
generalized Monte Carlo scheme, though it decreases exponentially
faster for
the standard Monte Carlo scheme.  The exponent $\eta$ becomes smaller as
$q$ is increased and the diffusion is possible even at low temperatures. 
This result can be anticipated based on data presented in figures 3
and 6.  This power-law temperature dependence of the diffusion constant
$D_{V}$ for $q\neq 1$ has been pointed out by Straub and
Andricioaei~\cite{SA2}.

\section{Concluding remarks}

In conclusion, we found that the so-called "quasi-ergodicity" 
in a double well potential encountered in the standard Metropolis 
Monte Carlo algorithm is largely circumvented by the
use of the generalized Monte Carlo algorithm of Andricioaei 
and Straub~\cite{AS2,AS3} based on Tsallis's~\cite{TS} generalized
statistical mechanics. Therefore this algorithm will be useful 
in the Monte Carlo study of any system, in particular, when
the first order phase transition occurs. We believe that
this algorithm will be useful, for example, in the study of 
the melting transition of clusters for which the J-walk
algorithm~\cite{FFD} and the multiple histrogram 
method~\cite{WA} have been utilized. It is also interesting
to apply this new algorithm to the problems where the slow
dynamics due to randomness or frustration is serious, for 
example, the spin-glass problems.  This generalized 
Monte Carlo algorithm based on Tsallis's statistical mechanics 
use the reweighting technique similar to the multicanonical 
method~\cite{BN}.  But this algorithm is much simpler than 
the multicanonical methods and appears to be more promising 
because it does not require an iterative determination of the
weighting factor by preliminary Monte Carlo runs.

Recently, a new formulation of the Tsallis statistics 
has been proposed~\cite{TMP}.  With a new choice of energy 
constraint, unfamiliar consequences in the previous formulation 
have been erased.  However, the Lagrange parameter used 
in the new formulation can be related to that in the previous 
formulation. Therefore, in performing Monte Carlo simulations 
with reweighting technique, we may use the previous formulation 
of the Tsallis statistics.


\begin{flushleft}
{\Large \bf Acknowlegement}
\end{flushleft}

Authors are grateful to the reviewer for several useful comments 
which were incorporated into the final version of the manuscript.
M. I. acknowledges support from the Hiroshima City University 
Grant for Special Academic Research. He is also grateful 
to the Department of Physics, Tokyo Metropolitan University,
where this work was initiated, for its hospitality to him.  


\newpage


\begin{figure}[th]
\epsfxsize=8cm \centerline{\epsfbox{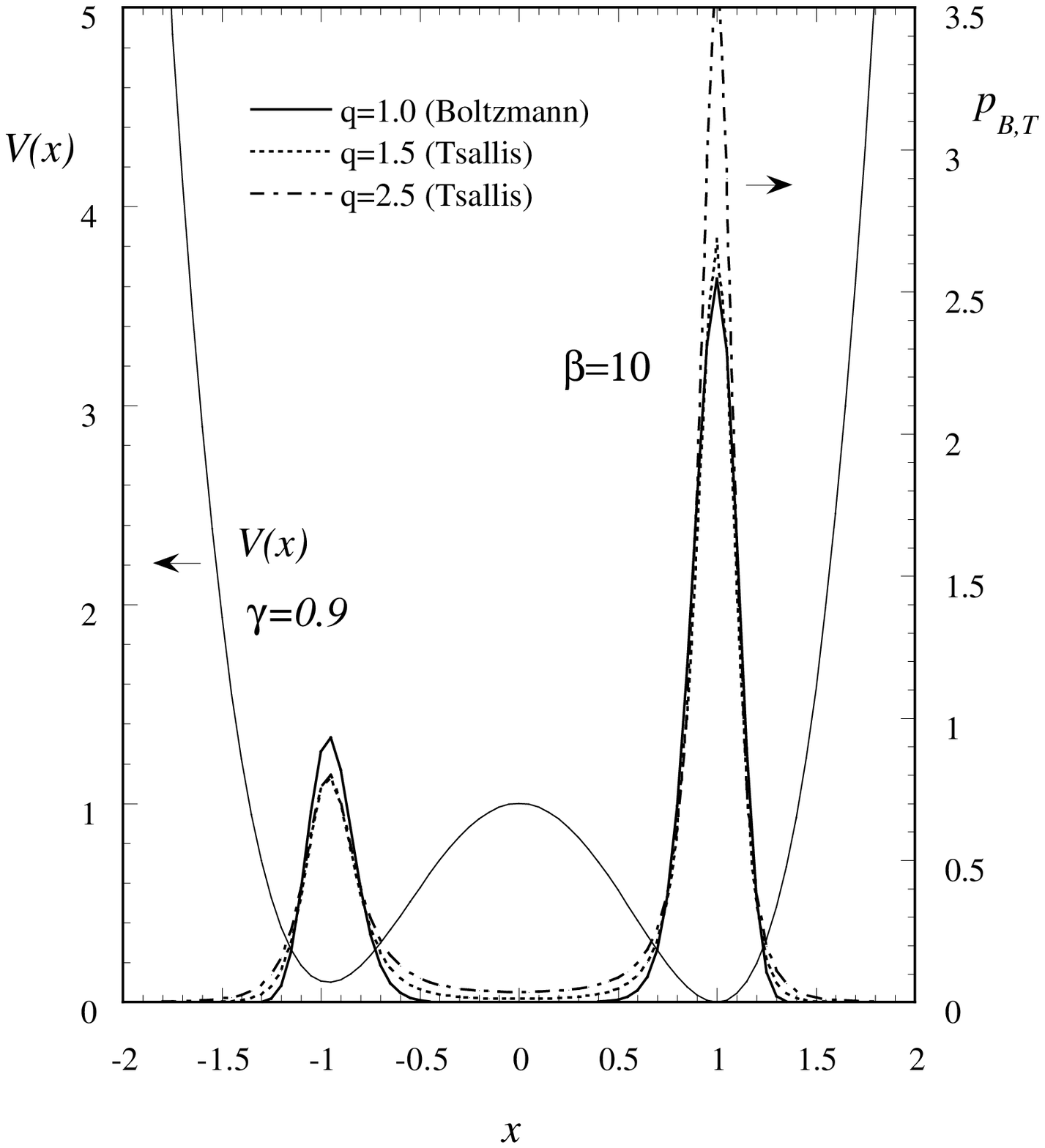}}
\caption{A classical double well 
potential (16) (thin solid curve),for $\gamma=0.9$, the Tsallis weight 
$p_{\rm T} \propto \exp(-\beta \overline{V}(x))$ with $q=1.5$ 
(dotted curve), $q=2.5$ (chain curve) and the Boltzmann 
weight $p_{\rm B}=\exp(-\beta V(x))/Z_{1}$ (thick solid curve) 
both at temperature $\beta=10$.  The Tsallis weights are 
renormalized.}
\label{fig:1}
\end{figure}

\begin{figure}[th]
\epsfxsize=8cm \centerline{\epsfbox{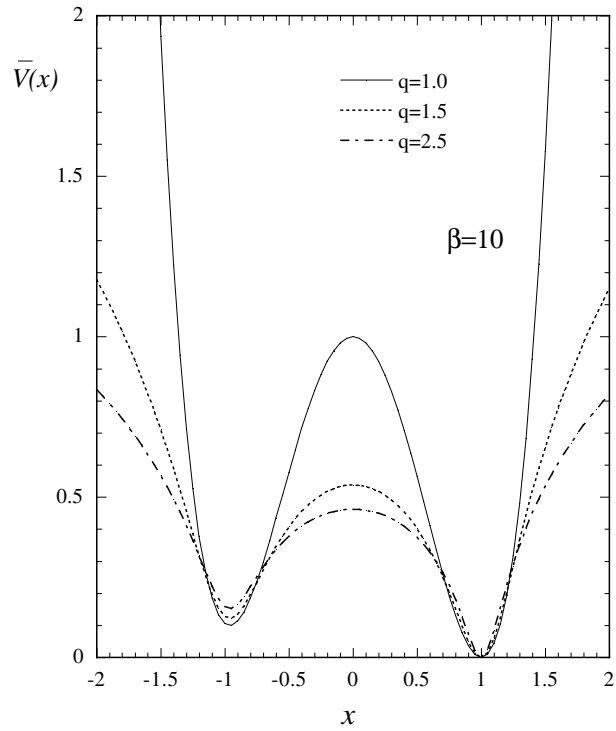}}
\caption{A smoothed potential landscape $\overline{V}(x)$ 
when $q=1.5$ (dotted curve) and $q=2.5$ (chain curve) for 
$\gamma=0.9$ and $\beta=10$, compared with the original 
potential landscape $V(x)$ ($q=1$, thin solid curve). 
The energy barrier is lowered by the smoothing defined 
by (3) of Tsallis's statistical mechanics.}
\label{fig:2}
\end{figure}

\begin{figure}[th]
\epsfxsize=8cm \centerline{\epsfbox{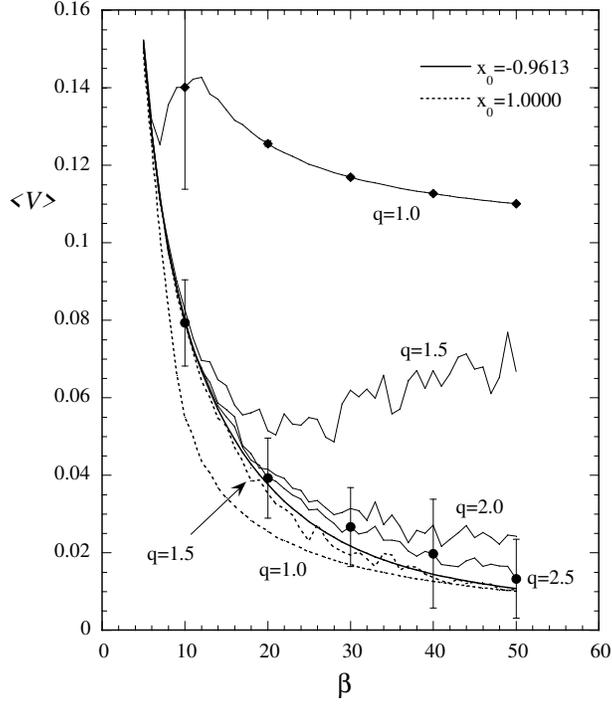}}
\caption{The classical average potential energy $<V>$ for 
the $\gamma=0.9$ potential as a function of $\beta$. The 
thin solid curves are initialized at the 
metastable $\alpha$-minimum $x_{0}=-0.9613$ 
with the standard Monte Carlo 
algorithm with the parameter $q=1$ and the generalized Monte 
Carlo algorithm with $q=1.5$, $q=2.0$ and $q=2.5$. 
The exact result is represented by the thick solid 
curve.  The results initialized at the global minimum $x_{0}=1$ 
and calculated from the Metropolis algorithm with $q=1$ and that from 
the generalized Monte Carlo algorithm with $q=1.5$ are shown 
by the dotted curve. All results are obtained from 100 independently 
initialized Monte Carlo walks, each consisting of 500 warm-up steps 
followed by $10^{4}$ steps with data accumulation.  Some 
representative error bars are 
also shown.}
\label{fig:3}
\end{figure}

\begin{figure}[th]
\epsfxsize=8cm \centerline{\epsfbox{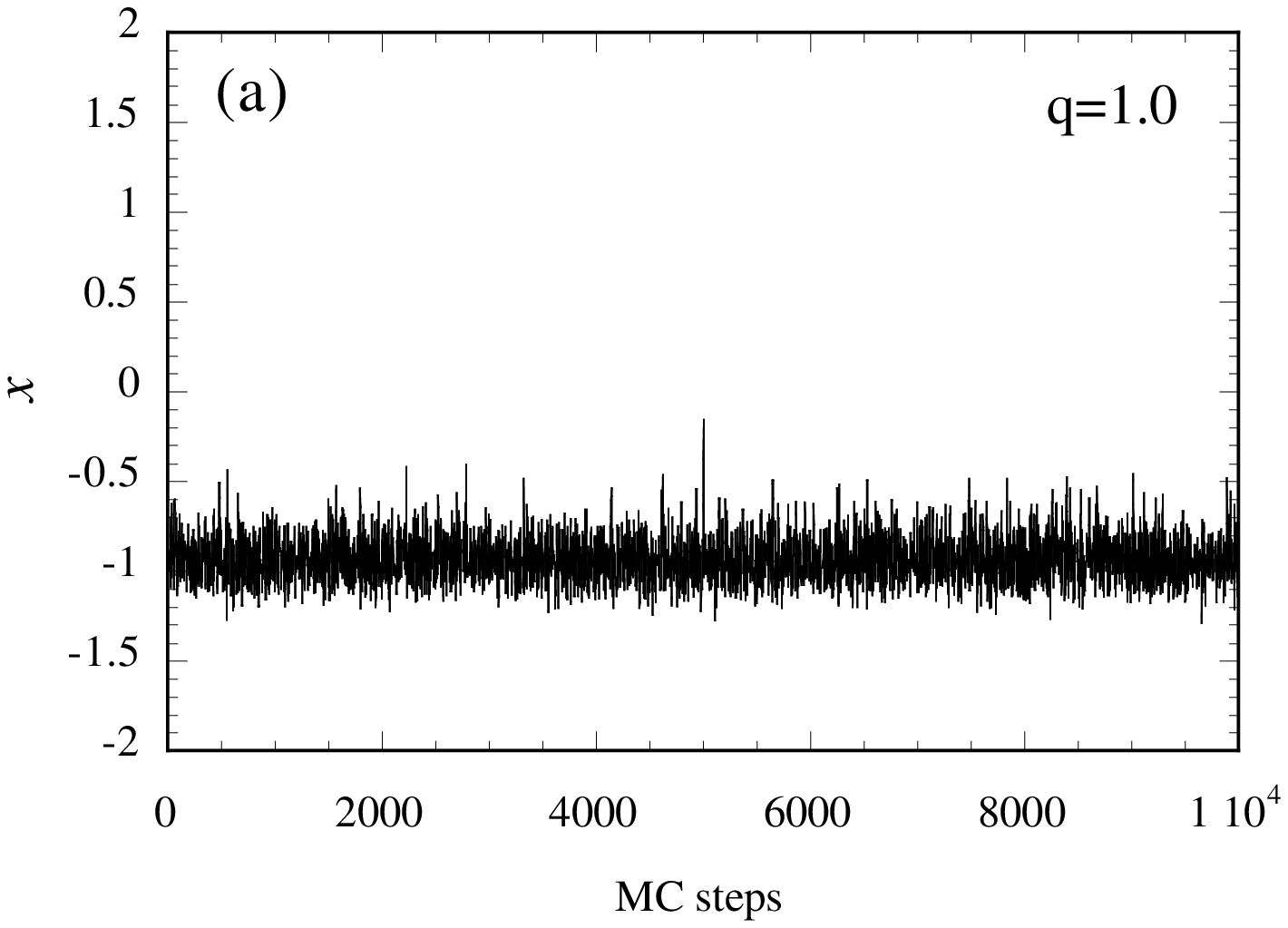}}
\vspace{4mm}
\epsfxsize=8cm \centerline{\epsfbox{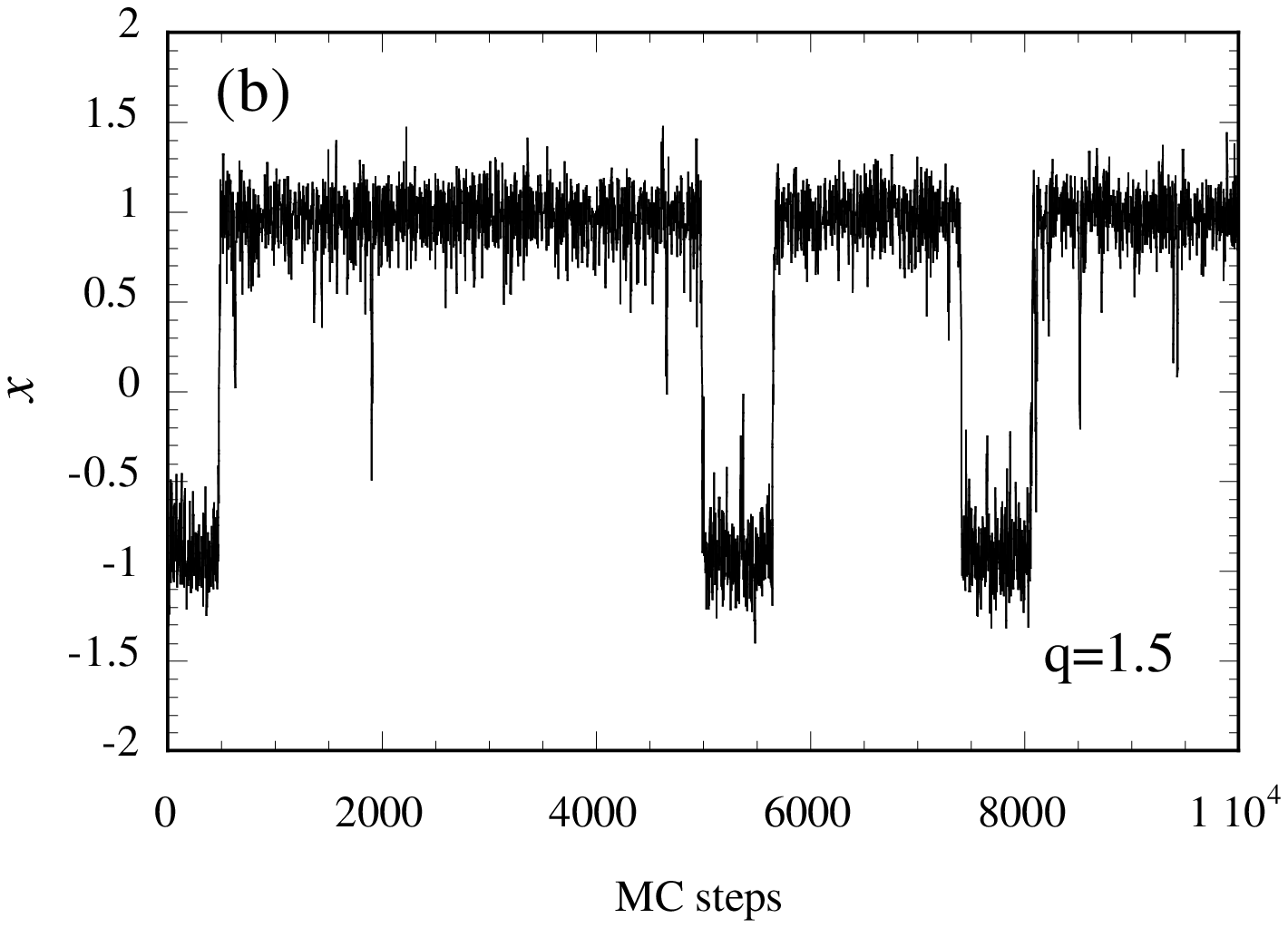}}
\vspace{4mm}
\epsfxsize=8cm \centerline{\epsfbox{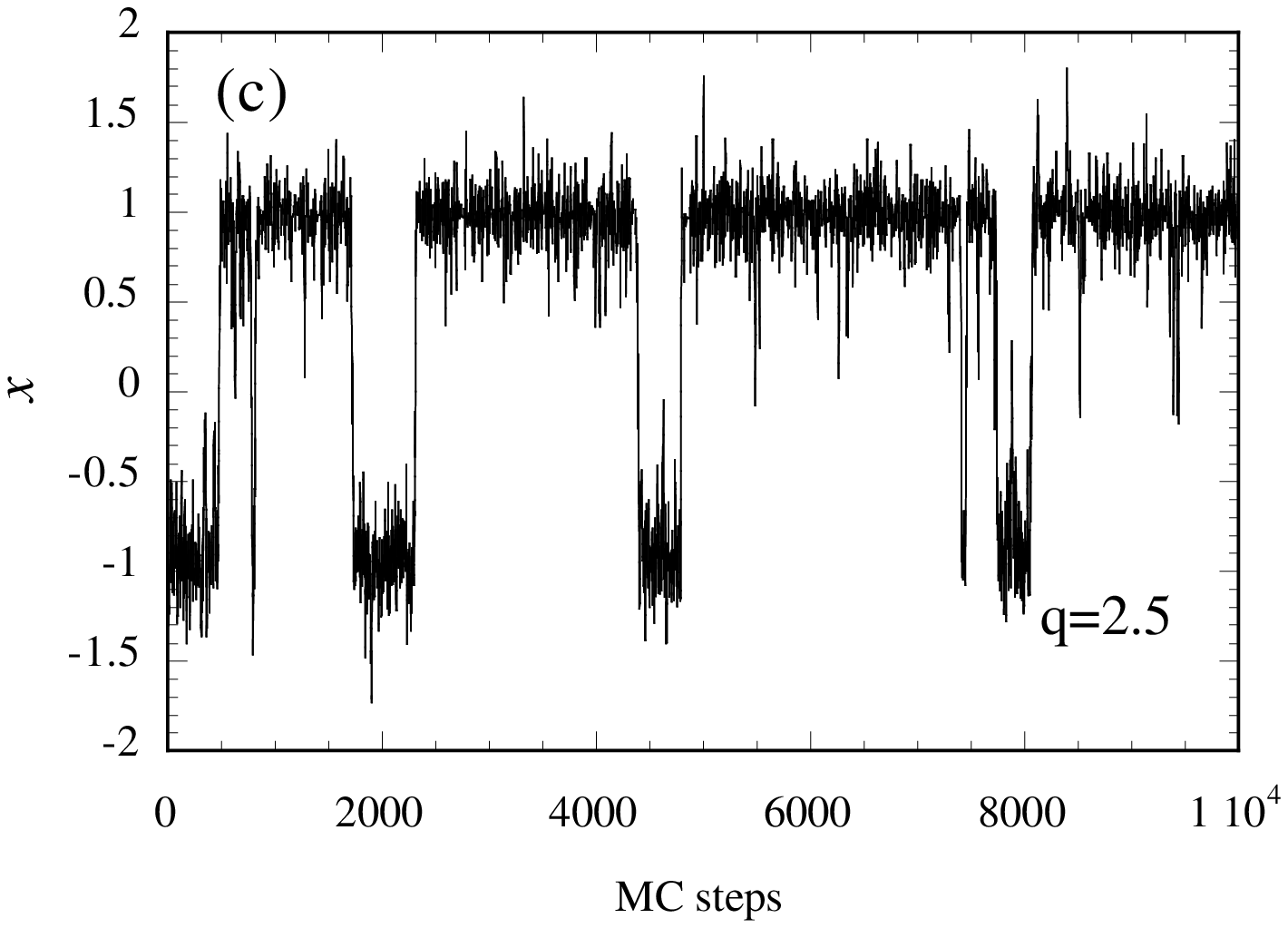}}
\caption{ Time series of the Monte Carlo walks generated by 
(a) the standard Monte Calro ($q=1$), (b) the Tsallis Monte 
Carlo with $q=1.5$, and (c) $q=2.5$ when $\beta=10$ and 
$\gamma=0.9$.  All walks start from the metastable 
$\alpha$-minimum, and 500 warm up steps are also included.  
The number of barrier crossing is greater
for larger $q$.} 
\label{fig:4}
\end{figure}

\begin{figure}[th]
\epsfxsize=8cm \centerline{\epsfbox{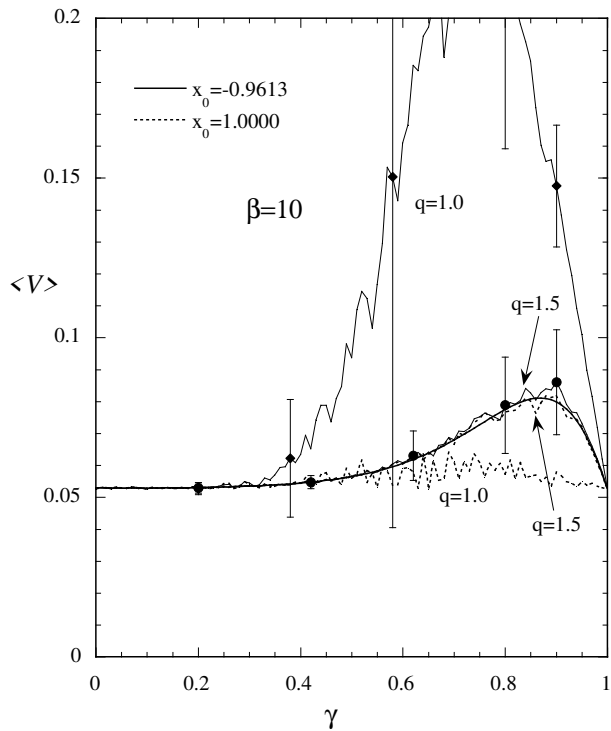}}
\caption{The average potential energy $<V>$ for the temperature 
$\beta=10$ as a function of the asymmetry parameter $\gamma$ of 
the double well potential. The solid and dotted curves are
used the same as in figure 3.  We only show the result and some 
representative error bars for the cases $q=1$ and $q=1.5$
of figure 3.} 
\label{fig:5}
\end{figure}

\begin{figure}[th]
   \def\halftext{.471\textwidth}
   \parbox{\halftext}{
      \epsfxsize=\halftext 
      \epsfbox{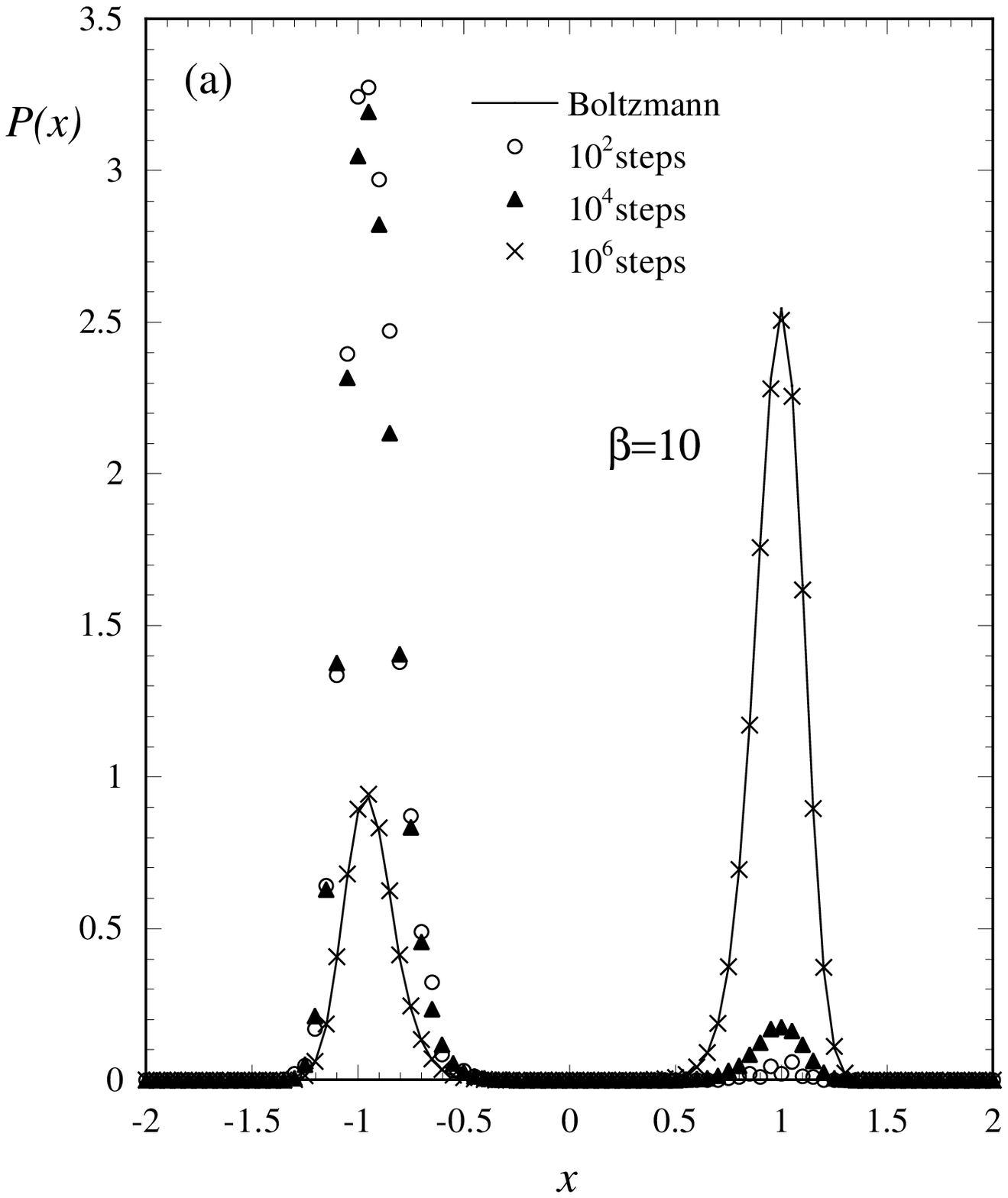}
   }
   \hspace{4mm}
   \parbox{\halftext}{
      \epsfxsize=\halftext 
      \epsfbox{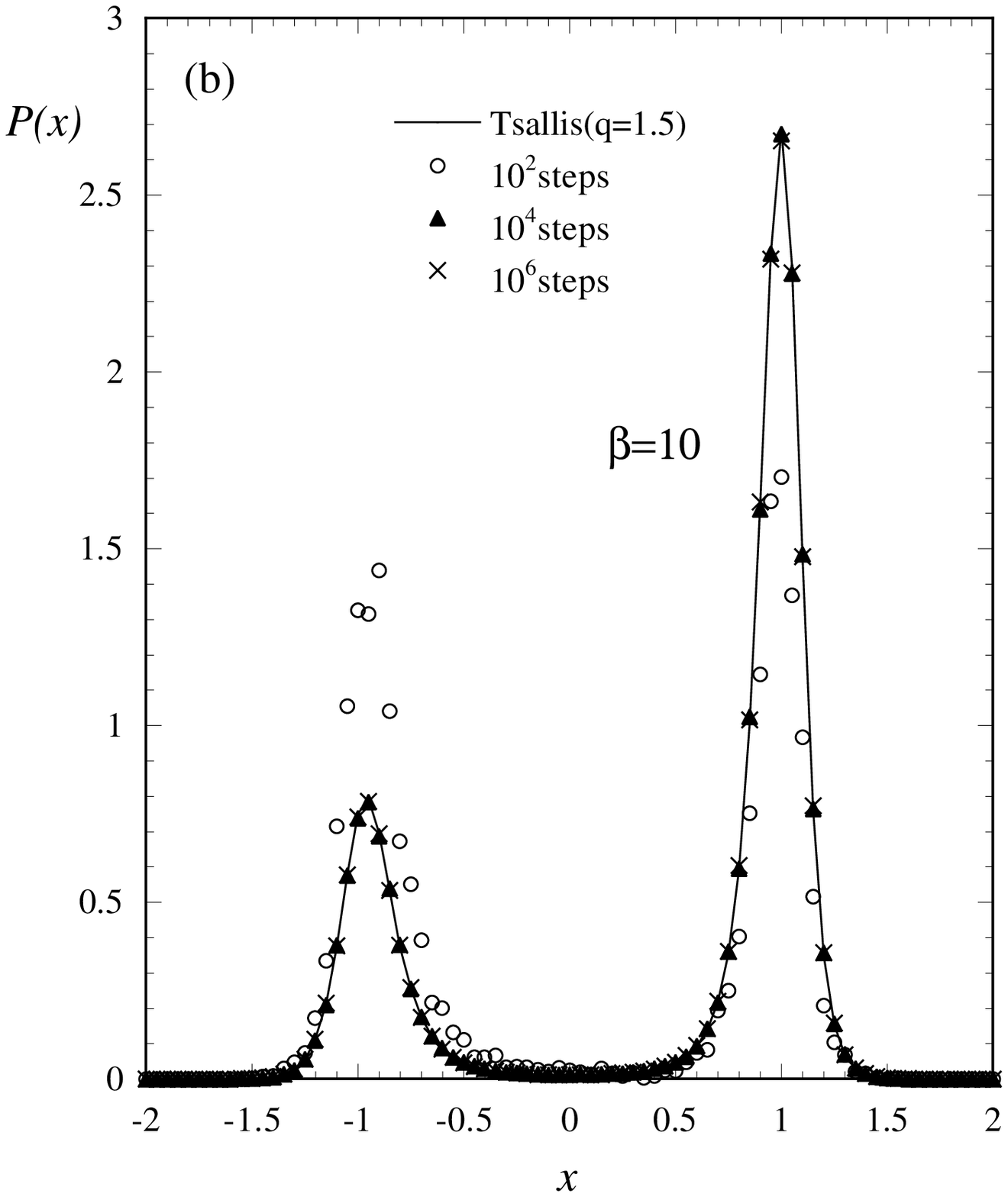}
   }
\caption{The Visiting probabilities $P(x)$ obtained from
(a) the standard Monte Carlo walk compared with the
Boltzmann weight, and (b) the generalized Monte Carlo 
walks compared with the Tsallis weight (c.f. figure 1).
Probability distribution $P(x)$ generated from the
generalized Monte Carlo scheme can recovers the 
equilibrium Tsallis weight within short Monte Carlo 
steps (a), while it cannot recover the equilibrium 
Boltzmann weight until rather long ($\sim 10^{6}$) Monte 
Carlo steps are executed (b).} 
\vspace*{20mm}
\label{fig:6}
\end{figure}

\begin{figure}[th]
\epsfxsize=8cm \centerline{\epsfbox{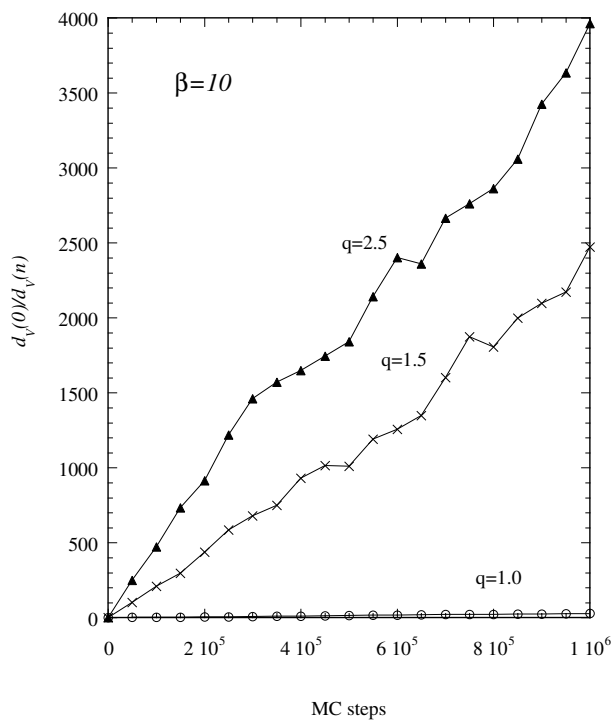}}
\caption{The normalized inverse of the generalized ergodic 
measure $d_{V}(0)/d_{V}(n)$ as a function of the number of
Monte Carlo steps $n$ calculated for $\gamma=0.9$ 
potential at $\beta=10$ using the standard Monte
Carlo scheme ($q=1$) and the generalized Monte Carlo
scheme ($q=1.5, 2.5$).  $d_{V}(0)/d_{V}(n)$ grows linearly
with the Monte Carlo steps $n$ and the diffusion is faster 
for larger $q$.} 
\label{fig:7}
\end{figure}

\begin{figure}[th]
   \def\halftext{.471\textwidth}
   \parbox{\halftext}{
      \epsfxsize=\halftext 
      \epsfbox{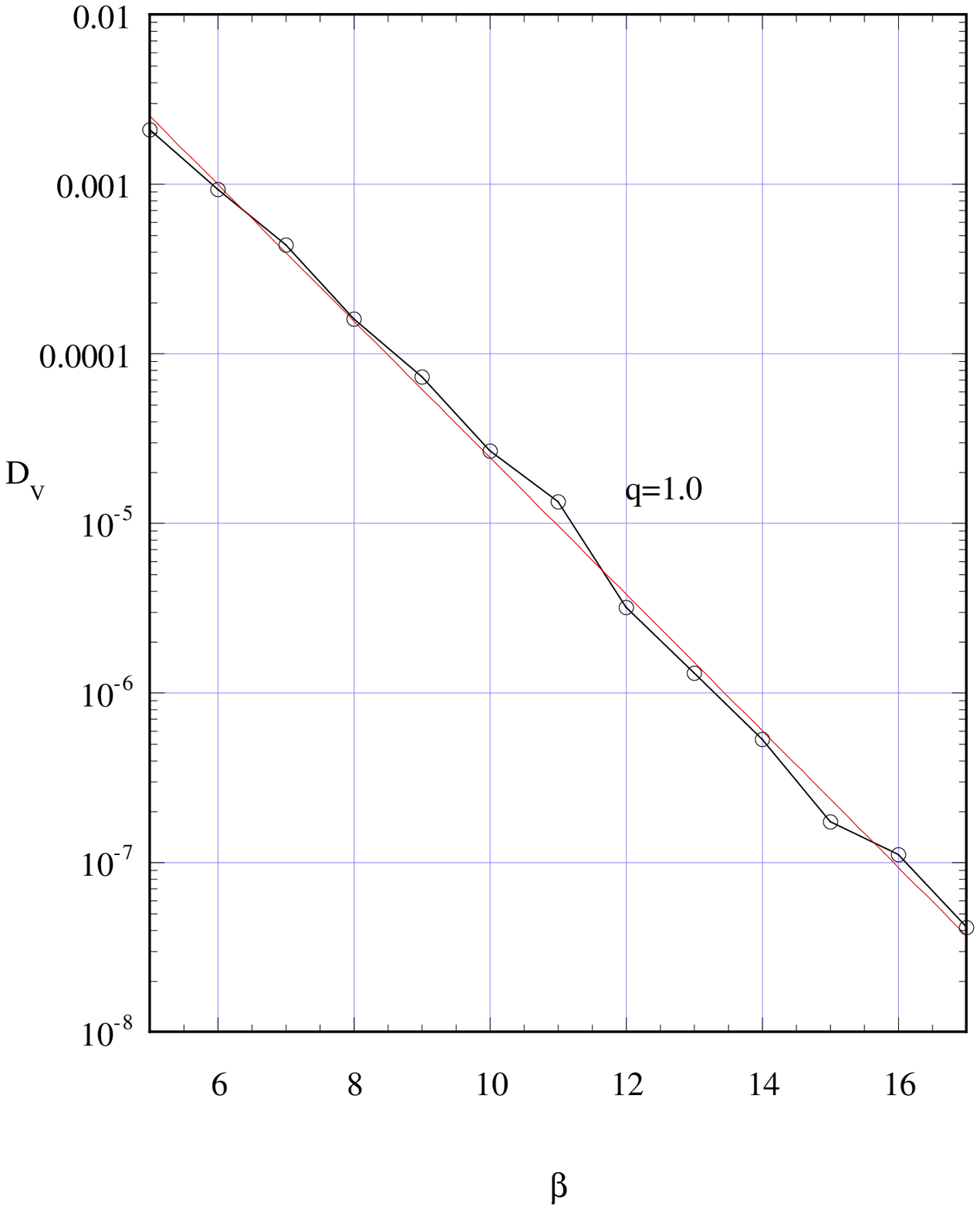}
   }
   \hspace{4mm}
   \parbox{\halftext}{
      \epsfxsize=\halftext 
      \epsfbox{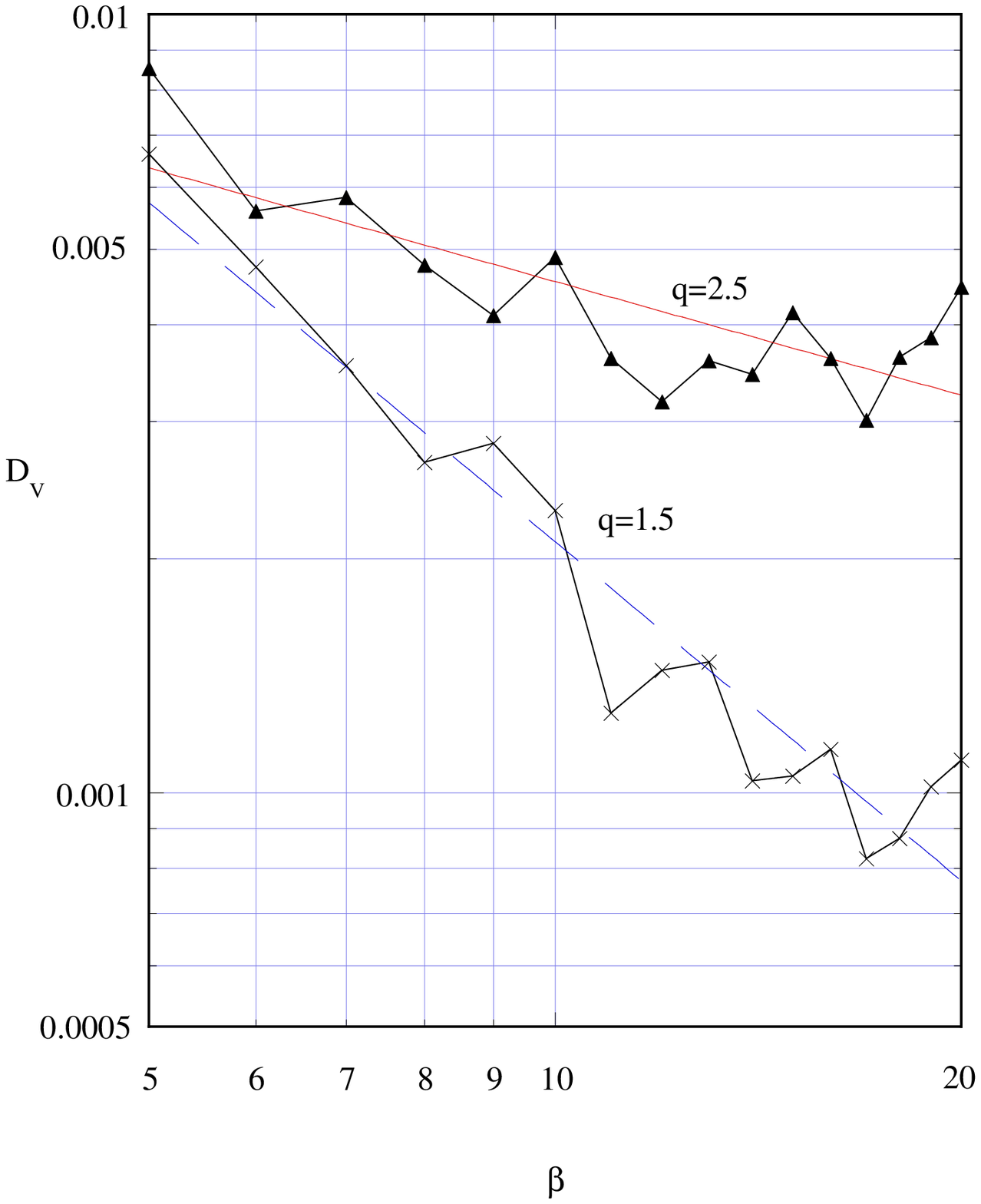}
   }
\caption{The diffusion constant $D_{V}$ as a function of 
temperature $\beta$.  (a) A straight line of the semi-log plot 
of $D_{V}$ vs $\beta$ for the standard Monte Carlo scheme 
indicates activation-type diffusion, while (b) straight 
lines of the log-log plot for the Tsallis Monte Carlo scheme 
in the high temperature region indicate slow power-law 
decrease of the diffusion constant as the temperature is lowered (larger
$\beta$).} 
\vspace*{20mm}
\label{fig:8}
\end{figure}

\newpage
\begin{table}[th]
\caption{The convergence of the Metropolis and the generalized 
Tsallis Monte Carlo algorithms at $\beta=10$. The average energy 
$<V>$, the standard deviation (STD) obtained from 100
independently initialized walks, each originating at the 
metastable $\alpha$-minimum.}

\vspace{4mm}
\begin{tabular}{c|c|c}
\hline
&Metropolis Monte Carlo & Tsallis Monte Carlo ($q=1.5)$ \\ 
steps & $<V>$ (STD) & $<V>$ (STD) \\
\hline
$10^{2}$ & $0.1500$ ($0.0177$) &$0.1065$ ($0.0518$) \\ 
$10^{3}$ & $0.1520$ ($0.0049$) &$0.0937$ ($0.0394$) \\ 
$10^{4}$ & $0.1475$ ($0.0192$) &$0.0798$ ($0.0162$) \\ 
$10^{5}$ & $0.1052$ ($0.0320$) &$0.0798$ ($0.0049$) \\ 
$10^{6}$ & $0.0805$ ($0.0124$) &$0.0800$ ($0.0017$) \\ 
Exact & $0.0799$	 & $0.0799$ \\
\hline
\end{tabular}
\end{table}

\end{document}